\documentclass[%
 reprint,
 amsmath,amssymb,
 aps,
pra,
]{revtex4-2}

\usepackage{graphicx}
\usepackage{dcolumn}
\usepackage{bm}
\usepackage{hyperref}
\usepackage[utf8]{inputenc}
\usepackage[T1]{fontenc}

\usepackage{xcolor}
\begin{document}


\title{Revisiting the displacement current: \\ Two key examples showing when and why it can be neglected}

\author{Álvaro Suárez$^1$, Arturo C. Martí$^{2}$, Martín Monteiro$^3$}

\affiliation{$^1$Departamento de Física, Consejo de Formación en Educación, Montevideo, Uruguay}
\affiliation{$^2$Instituto de F\'{i}sica,  Universidad de la Rep\'{u}blica, Uruguay}
\affiliation{$^3$Universidad ORT Uruguay, Montevideo, Uruguay}


\date{\today}

\begin{abstract}
This work explores the role of the displacement current in systems beyond capacitors, focusing on coaxial cables and resistors with alternating currents. 
Although its contribution, compared to that of the conduction current,
is negligible at low frequencies, the displacement current becomes essential for accurately describing electromagnetic fields in dynamic regimes. Introductory physics textbooks typically restrict the discussion of displacement currents to capacitor charging, which can limit the students’ understanding of its broader relevance and the interdependence of Maxwell’s equations. By analyzing two specific cases, we clarify the conditions under which the displacement current can be neglected, when simplified laws like Biot-Savart are valid, and where coupled electromagnetic equations are essential. This approach highlights the boundaries of common models and promotes a deeper understanding of dynamic field interactions.

\end{abstract}

\maketitle


\section{\label{Section 1}Introduction}

The displacement current plays a fundamental role in the formulation of Maxwell’s equations and is essential for a comprehensive understanding of electromagnetic phenomena. Historically, Maxwell introduced the displacement current in the 19th century as a modification to Ampère’s law, ensuring the internal consistency of Maxwell’s equations and paving the way for the prediction of electromagnetic waves \cite{berkson2014fields}. Moreover, from a modern theoretical standpoint, the displacement current also plays a crucial role in the covariant formulation of relativistic electrodynamics, guaranteeing the compatibility of Maxwell’s equations with the principle of charge conservation \cite{griffiths2013introduction}. Despite its conceptual significance, in widely recommended introductory physics textbooks \cite{suarez2023}, the displacement current is primarily addressed in the context of capacitor charging \cite{young2019university,serway2018physics,tipler2007physics,chabay2015matter,giancoli2014physics,walker2014halliday}, resulting in a limited treatment that can influence students’ conceptual understanding \cite{suarez2024learning}.

This limited approach may lead students to believe that the displacement current is relevant exclusively within capacitor-related problems. Moreover, in studies involving the electric and magnetic fields generated in coaxial cables and resistors with time-varying currents, the contribution of the displacement current is often omitted without providing a clear explanation for such exclusion. One possible reason is that a formal treatment considering the displacement current to determine the fields in these systems exceeds the scope of introductory courses. However, this incomplete presentation not only risks leading students to assume that the displacement current should never be considered, but also to mistakenly believe that Maxwell’s equations can be applied independently in problems involving dynamic fields, when in fact they form a coupled system of equations \cite{Allred,suarez2022}.

Recent physics education research has documented several specific difficulties that students encounter concerning displacement current and Ampère–Maxwell’s law. In particular, students show persistent difficulty recognizing the displacement current concept beyond the standard capacitor charging example \cite{suarez2024learning}. Additionally, they often struggle to understand that a time-varying electric field necessarily implies the existence of an associated magnetic field \cite{suarez2024learning}. Students also commonly exhibit interference between the concepts involved in Faraday’s and Ampère–Maxwell’s laws \cite{suarez2024learning}, as well as significant difficulty adequately interpreting magnetic field circulation within the framework of Ampère–Maxwell’s law \cite{Suarez2025IOP}. These findings emphasize the need for instructional strategies explicitly designed to address these conceptual challenges.

In a previous paper \cite{suarez2025displacement}, we explicitly addressed these instructional challenges by presenting unconventional examples appropriate for introductory university-level physics instruction. In that work, we analyzed these examples in detail and provided general suggestions about how instructors might utilize them to address common student difficulties related to displacement current and Ampère–Maxwell’s law. The goal of that previous effort was to offer physics instructors insights into possible teaching approaches that can support students in developing a more integrated understanding of electromagnetic concepts.

The present article expands and complements that earlier analysis by examining realistic scenarios involving alternating currents in coaxial cables and resistors, which are typically absent from introductory discus
sions. Specifically, we analyze coaxial cables and resistors carrying alternating currents to clearly establish conditions under which displacement current contributions are negligible compared to conduction currents. In these cases, simplified laws such as Biot–Savart or the magnetostatic form of Ampère’s law remain valid approximations. Conversely, when displacement current is significant, a full treatment using Maxwell’s coupled equations is required. By clarifying these conditions, we explicitly identify the limits of validity for simplified electromagnetic models typically applied to real systems.

\section{\label{resistor}Displacement current in a resistor} 

The analysis of displacement current in resistors reveals that this phenomenon is not confined to capacitors but arises in any situation where the electric field varies with time, even within a conductor. This case not only underscores the universality of displacement current but also highlights key differences between conduction current density, which is proportional to the electric field, and displacement current density, which depends on its temporal variation. Furthermore, it allows for an exploration of the conditions under which displacement current is negligible compared to conduction current, thereby defining the limits of validity for using the Biot-Savart law to calculate magnetic fields.

In static or slowly varying systems, where electric fields exhibit negligible time dependence, the magnetic field can be accurately described by the magnetostatic form of Ampère’s law:
\begin{equation}
\nabla \times \mathbf{B} = \mu_0 \mathbf{J}_c,
\end{equation}
where $\mu_0$ is the vacuum permeability and $\mathbf{J}_c$ the conduction current density. To obtain the magnetic field explicitly, it is necessary to combine this equation with Gauss’s law for magnetism, $\nabla \cdot \mathbf{B} = 0$,  which expresses the solenoidal nature of the magnetic field. This condition allows the introduction of the magnetic vector potential $\mathbf{A}$, such that $\mathbf{B} = \nabla \times \mathbf{A}$. Under the Coulomb gauge condition, $\nabla \cdot \mathbf{A} = 0$, one obtains the Poisson equation for the vector potential, whose solution leads to the Biot–Savart law \cite{feynman1963feynman}:
\begin{equation}
\label{Biot corrientes}
\mathbf{B}=\frac{\mu_{0}}{4 \pi} \int \frac{\mathbf{J}_{c}\left(r^{\prime}\right)}{|\mathbf{R}|^{3}} \times \mathbf{R} d v^{\prime}
\end{equation}
which expresses the magnetic field at point $\mathbf{R}$ in terms of the
current distribution $\mathbf{J}_c(\mathbf{r}')$. This law is not restricted to static cases, but its use as an approximation in time-dependent contexts assumes that the currents vary slowly enough so that their time derivatives and the associated radiation and delay effects can be neglected. In that sense, it reflects the application of Ampère's law in a regime where the displacement current plays no significant role.

In general, however, electric fields vary with time and displacement currents cannot be neglected. In such dynamic situations, the magnetostatic approximation is no longer valid, and Ampère’s law must be extended to include the effects of time-dependent electric fields. This extension, known as  Ampère–Maxwell law, introduces the displacement current  to ensure consistency with charge conservation
\begin{equation}
\nabla \times \mathbf{B} = \mu_0 \mathbf{J}_c + \mu_0 \varepsilon_0 \frac{\partial \mathbf{E}}{\partial t},
\end{equation}
where, $\varepsilon_0$ is the vacuum permittivity and $E$ is the elefctric field.
When this equation is considered in conjunction with the full set of Maxwell’s equations, it becomes evident that electric and magnetic fields are intrinsically coupled and must be determined simultaneously.

\begin{figure}
\centerline{\includegraphics[width = 0.9\columnwidth]{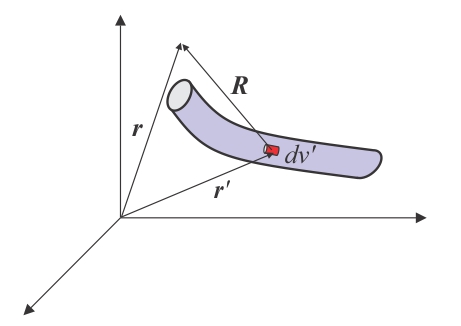}}
\caption{Small section of a conductor carrying a conduction current.}
\label{Fig campos y biot}
\end{figure}

The general solution to this coupled system in vacuum leads to the retarded potentials and, ultimately, to Jefimenko’s equations, which explicitly describe how time-varying sources give rise to electromagnetic fields \cite{griffiths2013introduction}. In particular, the magnetic field generated by a volume element $dv^{\prime}$ located at a distance $R$, as illustrated in Fig. \ref{Fig campos y biot}, is given by \cite{jefimenko1966electricity}
\begin{equation}
\label{B jefimenko}
\mathbf{B}=\frac{\mu_{0}}{4 \pi} \int\left(\frac{\mathbf{J}_c\left(r^{\prime}, t_{r}\right)}{\left|\mathbf{R}\right|^{3}}+\frac{\dot{\mathbf{J}}_c\left(r^{\prime}, t_{r}\right)}{c\left|\mathbf{R}\right|^{2}}\right) \times\mathbf{R} d v^{\prime} 
\end{equation}
where, $t_{r}=t-R / c$ is the delayed time, which reflects the finite speed at which electromagnetic interactions propagate through space. This expression makes explicit that the magnetic field depends not only on the instantaneous current density but also on its time derivative. The second term arises directly from the inclusion of the displacement current in the Ampère-Maxwell law and is essential to capture the physical effects associated with sources that vary over time, particularly those related to electromagnetic radiation.

These considerations can now be applied to a system of practical relevance. To illustrate the ideas above, let us consider a resistor of resistivity $\rho$ and electric permittivity $\varepsilon$, subjected to an axial time-varying electric field of the form
\begin{equation}
E(t)=E_{0} \sin (\omega t) 
\end{equation}
The  density of the conduction current along the resistance is given by
\begin{equation}
\label{Jc resistor}
J_{c}(t)=\frac{E(t)}{\rho}=\frac{E_{0} \sin (\omega t)}{\rho}
\end{equation}
and, the displacement current density is given by
\begin{equation}
\label{Jd resistor}
J_{D}(t)=\varepsilon \dot{E}=\varepsilon \omega E_{0} \cos (\omega t). 
\end{equation}
Since the current changes with time (and consequently, a displacement current exists in the resistor), the magnetic field must be calculated using Jefimenko's Eq.~\ref{B jefimenko}.

Now, let us establish the conditions under which Biot-Savart’s law (Eq. \ref{Biot corrientes}) remains a valid approximation in situations where the conduction current changes with time. We expand $\mathbf{J}_{c}\left(r^{\prime}, t_{r}\right)$ in a Taylor series around $t$
\begin{equation}
\label{desarrollo Jc}
\mathbf{J}_{c}\left(r^{\prime}, t_{r}\right)=\mathbf{J}_{c}(t)+\left(t_{r}-t\right) \dot{\mathbf{J}}_{c}\left(r^{\prime}, t\right)+\frac{\left(t_{r}-t\right)^{2}}{2} \ddot{\mathbf{J}}_{c}\left(r^{\prime}, t\right)+\cdots,
\end{equation}
as $t_{r}=t-R / c$, it results
\begin{equation}
\label{desarrollo Jc tr}
\mathbf{J}_{c}\left(r^{\prime}, t_{r}\right)=\mathbf{J}_{c}(t)-\frac{R}{c} \dot{\mathbf{J}}_{c}\left(r^{\prime}, t\right)+\frac{R^{2}}{2 c^{2}} \ddot{\mathbf{J}}_{c}\left(r^{\prime}, t\right)+\cdots
\end{equation}

Assuming that the current density changes with time ‘slowly enough’, we can neglect high-order terms. Substituting Eq.~\ref{desarrollo Jc tr} in Eq.~\ref{B jefimenko} 
it is easy to obtain
\begin{equation}
\label{B con desarrollo}
\mathbf{B}=\frac{\mu_{0}}{4 \pi} \int \frac{1}{R^{3}}\left[\mathbf{J}_{c}(t)-\frac{R}{c} \dot{\mathbf{J}}_{c}\left(r^{\prime}, t\right)+\frac{R}{c} \dot{\mathbf{J}}_{c}\left(r^{\prime}, t_{r}\right)\right] \times \mathbf{R} d v^{\prime}
\end{equation}
Now developing $\dot{\mathbf{J}}_{c}\left(r^{\prime}, t_{r}\right)$ and keeping the
leading term, $\dot{\mathbf{J}}_{c}\left(r^{\prime}, t_{r}\right)=\dot{\mathbf{J}}_{c}\left(r^{\prime}, t\right)
$
we substitute in  Eq.~\ref{B con desarrollo}, to find that when the
current slowly varies in time the magnetic field is given by
\begin{equation}
\label{B lentamente}
\mathbf{B}_{J \text {slow}}=\frac{\mu_{0}}{4 \pi} \int \frac{\mathbf{J}_{c}\left(r^{\prime}, t\right)}{R^{3}} \times \mathbf{R} d v^{\prime}
\end{equation}
which is precisely  Biot-Savart's law with the current density evaluated at time \textit{t}.

We note that, to first order, Biot-Savart's law is still a very good approximation for the magnetic field generated by a time-varying current as long as the variation is slow \cite{griffiths2013introduction}.
By developing in series the  current density and truncating the development in the second term, we are neglecting the higher order terms. Therefore, the
slow variation of the current  implies that the third term  is much smaller than the second term,
\begin{equation}
\label{comparación derivadas J}
\frac{R^{2}}{2 c^{2}} \ddot{\mathbf{J}}_{c} \ll \frac{R}{c} \dot{\mathbf{J}}_{c}
\end{equation}

To appreciate the physical significance of this relationship, let us note that for a sinusoidally varying conduction current density, as given by Eq.~\ref{Jc resistor}, the time derivative is proportional to the frequency $\omega$ and its second derivative to $\omega^{2}$. Taking this aspect into account and simplifying Eq.~\ref{comparación derivadas J}, we find that when the current varies slowly

\begin{equation}
\label{comparación derivadas 2}
\frac{R\omega }c \ll 1. 
\end{equation}

Let us denote $\Delta t= R / c$ the time needed for the magnetic field  to arrive from the source to the point of interest and the period as $T= 2 \pi / \omega $. The condition for assuming that the current varies slowly is

\begin{equation}
\label{comparación derivadas 3}
\Delta t \ll T 
\end{equation}

We see then that when the time it takes for the magnetic field to propagate from its source to the place where it is measured is much less than the time it takes for the current to change, the Biot-Savart's law is  still a very good approximation.

Since Biot–Savart’s law follows from the magnetostatic approximation of Maxwell’s equations, the condition given by Eq.~\ref{comparación derivadas J} is equivalent to stating that, under these circumstances, the magnetic field can be determined using Maxwell’s equations in a decoupled manner. Consequently, the displacement current term, which couples Ampère-Maxwell’s law with Faraday’s law, can be neglected.

Although the analysis carried out is beyond the scope of an introductory course in electromagnetism, the discussion of the qualitative point of view as to why it seems reasonable to use Biot-Savart's law when currents vary slowly can be carried out in introductory courses. Thus, the study of the displacement current in a resistor allows us not only to recognise that it can manifest itself in other elements of a circuit besides the capacitor, but also to discuss the framework of validity of Biot-Savart's law.

\section{\label{section 2}Displacement current in a coaxial cable}

The analysis of electromagnetic fields in a coaxial cable is useful in demonstrating where the limits of Ampere's law lie. The coaxial cable is a well-known system and is widely developed in textbooks in two extreme cases. One is the well-known case of a coaxial cable with direct current, in which a constant magnetic field is generated that is confined to the interior of the coaxial cable. The other well-studied case is that of high-frequency alternating current, which gives rise to what is known as a waveguide or transmission line, in which stationary electromagnetic waves are produced. What is meant by high frequencies and what happens to the displacement current and the Ampère-Maxwell law in the case of intermediate frequencies is analyzed and demonstrated in this example.

\begin{figure}
\centering{\includegraphics[width = 0.85\columnwidth]{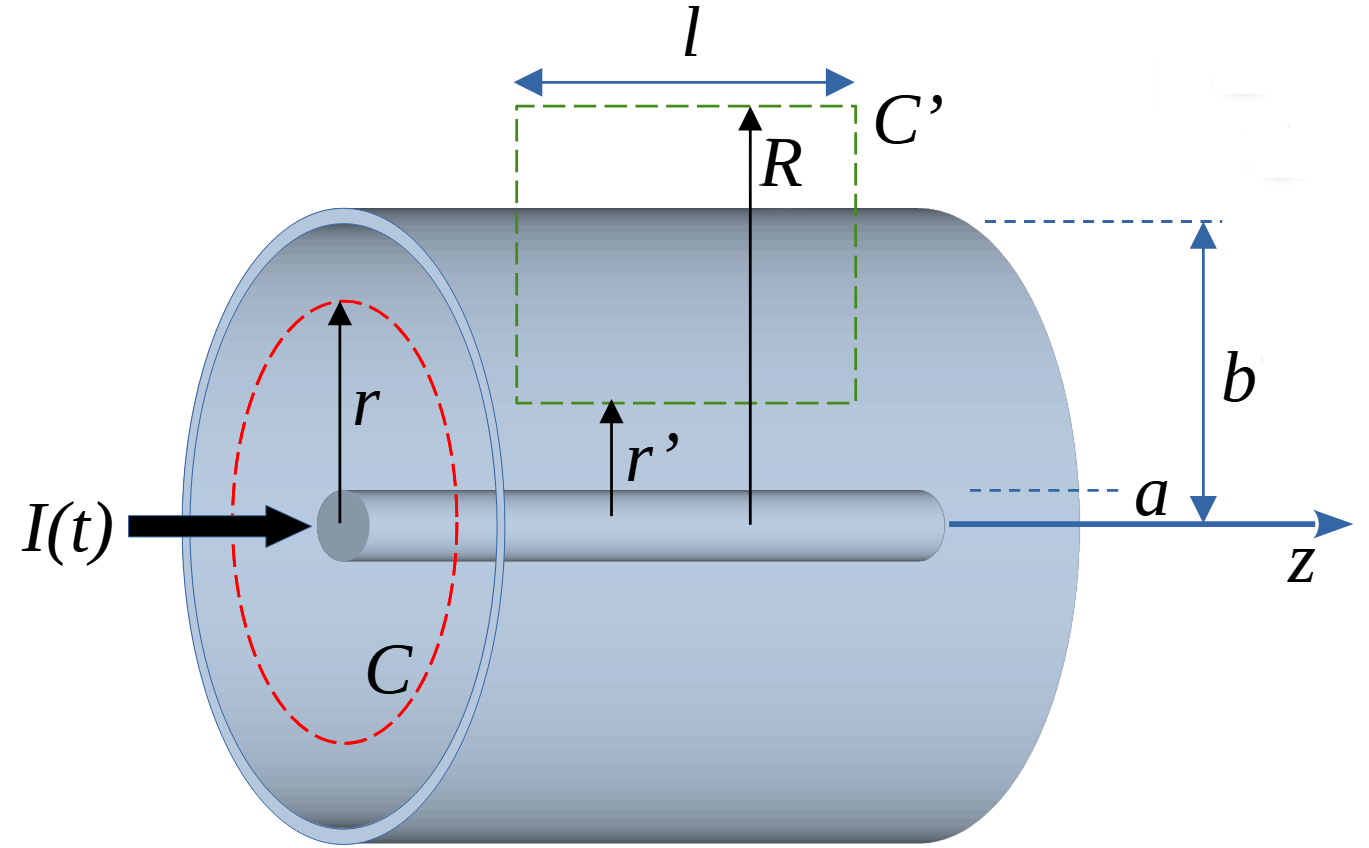}}
\centering{\includegraphics[width = 0.95\columnwidth]{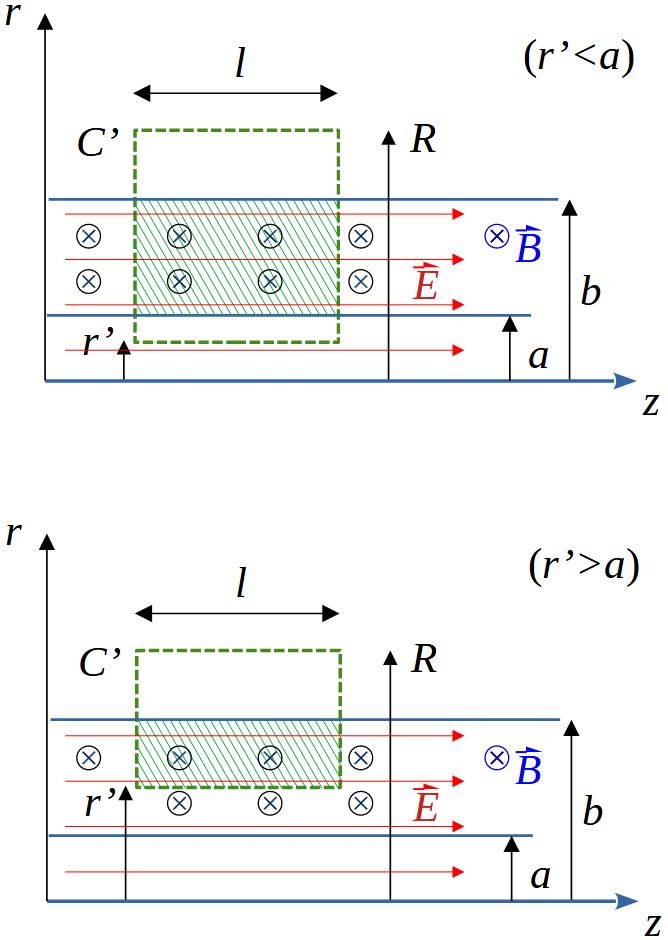}}
\caption{
Top panel: a longitudinal section of the coaxial and the curves $C$ and ${C}^{\prime}$ considered in the analysis. 
Middle panel: if ${r}^{\prime}<a$ the magnetic flux is produced by the field Eq.~\ref{B2 coaxial AM} between $a$ and $b$. The azimuthal magnetic field and the longitudinal electric field are also shown.
Lower panel: if ${r}^{\prime}>a$ the magnetic flux is produced by the field 
Eq.~\ref{B2 coaxial AM}  between ${r}^{\prime}$  and $b$.}
\label{Figcoaxil}
\end{figure}

Let us consider a coaxial cable consisting of two concentric, thin-walled, hollow conducting cylinders: an inner conductor of radius $a$, carrying a uniform time-dependent current 
$I(t)=I_{0} \cos (\omega t)$, and an outer conductor of radius $b$, through which the return current flows. This current generates electromagnetic fields in the insulating region that separates both conductors. To simplify the analysis, without losing validity in the final conclusions, let us consider that the conductors are ideal, with zero resistivity $\rho \approx 0$, and that the permittivity and permeability of the insulator are $\varepsilon_{0}$ and $\mu_{0}$. Applying Ampère's law on a circumference $C$ with radius $r$ as shown in Fig. \ref{Figcoaxil} , upper panel, and assuming that the current is quasi-steady, meaning that it varies very slowly (a concept we will clarify later), the magnetic field is in the azimuthal direction and is confined to the insulating region between a and b. Its magnitude in this region results in
\begin{equation}
\label{B2 coaxial AM}
\mathrm{B}=\frac{\mu_{0}}{2 \pi} I_{0} \cos (\omega t) \frac{1}{r} \quad(a<r<b).
\end{equation}

This time-dependent azimuthal magnetic field implies the existence of a longitudinal electric field that can be obtained by means of Faraday's law. 

Let us consider a rectangular shaped curve $\mathrm{C}^{\prime}$ with two of its sides parallel to the longitudinal direction at distances $r^{\prime}$ and $R$ from the $z$-axis (as shown in the middle and bottom panel of Fig.~\ref{Figcoaxil}), then the magnetic flux inside $\mathrm{C}^{\prime}$ is given by
\begin{equation}
\label{flujo1}
\Phi_{B}=\int \mathbf{B} \cdot \mathbf{d} \mathbf{A}=\int_{0}^{l} \int_{r^{\prime}}^{R} B d r d z
\end{equation}
If ${r}^{\prime}<a$, the magnetic flux results
\begin{equation}
\label{flujo2}
\Phi_{B}=\int_{0}^{l} \int_{a}^{b} \frac{\mu_{0}}{2 \pi} I_{0} \cos (\omega t) \frac{1}{r} d r d z
\end{equation}
which can readily integrated to obtain
\begin{equation}
\label{flujo4}
\Phi_{B}=\frac{\mu_{0}}{2 \pi} I_{0} \cos (\omega \mathrm{t}) l\operatorname{Ln}\left(\frac{b}{a}\right), \quad\left(0 \leq r^{\prime} \leq a\right).
\end{equation}

In the insulating region, $a<r^{\prime}<b$, the magnetic flux can be expressed as
\begin{equation}
\label{flujo5}
\Phi_{B}=\int_{0}^{l} \int_{r^{\prime}}^{b} \frac{\mu_{0}}{2 \pi} I_{0} \cos (\omega t) \frac{1}{r} d r d z, 
\end{equation}
integrating this expression we obtain
\begin{equation}
\label{flujo7}
\Phi_{B}=\frac{\mu_{0}}{2 \pi} I_{0} \cos (\omega t) l \operatorname{Ln}\left(\frac{b}{r^{\prime}}\right), \quad\left(a \leq r^{\prime} \leq b\right).
\end{equation}

The electric field in the inner region can be obtained from Faraday's law considering the curve $\mathrm{C}^{\prime}$ 
\begin{equation}
\label{faraday}
\oint_{C^{\prime}} \mathbf{E} \cdot \mathbf{d} \mathbf{l}=-\dot{\Phi}_{B},
\end{equation}
and nothing that an azimuthally varying magnetic field in a coaxial structure does not generate an electric field outside. This follows from the fact that the displacement current (and any associated charge redistribution) is confined within the coaxial region, and the outer conductor acts as a shield due to its role in enforcing boundary conditions that prevent external fields. Furthermore, from Maxwell’s equations, the induced electric field associated with an azimuthal magnetic field must have either a radial or a longitudinal component. In a coaxial system, a radial electric field would require a charge imbalance or discontinuities that are not present in this configuration. Consequently, the only physically consistent component of the induced electric field is longitudinal.

Let us substitute Eq.~\ref{flujo4} in Eq.~\ref{faraday}, resulting 
\begin{equation}
\label{E1}
E l=\omega \frac{\mu_{0}}{2 \pi} I_{0} \sin (\omega t) l \operatorname{Ln}\left(\frac{b}{a}\right),
\quad(0 \leq r \leq a).
\end{equation}

The electric field in the insulating region can be similarly obtained using Eq.~\ref{flujo7} and  Eq.~\ref{faraday},
\begin{equation}
\label{E2}
E l=\omega \frac{\mu_{0}}{2 \pi} I_{0} \sin (\omega t) l \operatorname{Ln}\left(\frac{b}{r}\right), \quad(a<r<b).
\end{equation}
Finally, the previous expressions  can be easily rearranged to give
\begin{equation}
\label{E3}
E=\frac{\mu_{0}}{2 \pi} \omega I_{0} \sin (\omega t)\ln \left(\frac{b}{a}\right), \quad(0 \leq r \leq a)
\end{equation}
and 
\begin{equation}
\label{E4}
E=\frac{\mu_{0}}{2 \pi} \omega I_{0} \sin (\omega t) \ln \left(\frac{b}{r}\right), \quad(a<r<R).
\end{equation}

This time-dependent electric field gives rise to a displacement current, which, in principle, prevents the direct application of Ampère’s law. Solving for the complete electromagnetic field requires a full solution to Maxwell’s equations, which is beyond the scope of introductory courses. However, it is useful to determine the conditions under which the displacement current can be neglected, allowing Ampère’s law to remain a valid approximation. The displacement current associated with the time-dependent electric field inside a cylindrical region of radius 
$r$, ( $a<r<b, a \approx$ 0 ), is given by:
\begin{equation}
\label{Id1}
\begin{split}
I_{D}=\varepsilon_{0} \int_{S} \frac{d \mathbf{E}}{d t} \cdot \mathbf{d A} \approx \\
\approx \frac{1}{4} \varepsilon_{0} \mu_{0} \omega^{2} I_{0} \cos (\omega t) r^{2}\left[2 \ln \left(\frac{b}{r}\right)+1\right].
\end{split}
\end{equation}

Note that the electric field inside the conductor, which is associated with the conduction current, does not contribute to the displacement current. This is because the field is negligible due to the conductor's extremely low resistivity $\rho \approx 0$, as mentioned earlier. Consequently, since $E = \rho J \approx 0$, the displacement current within the conductor can be ignored.

\begin{figure}
\centerline{\includegraphics[width = 0.99\columnwidth]{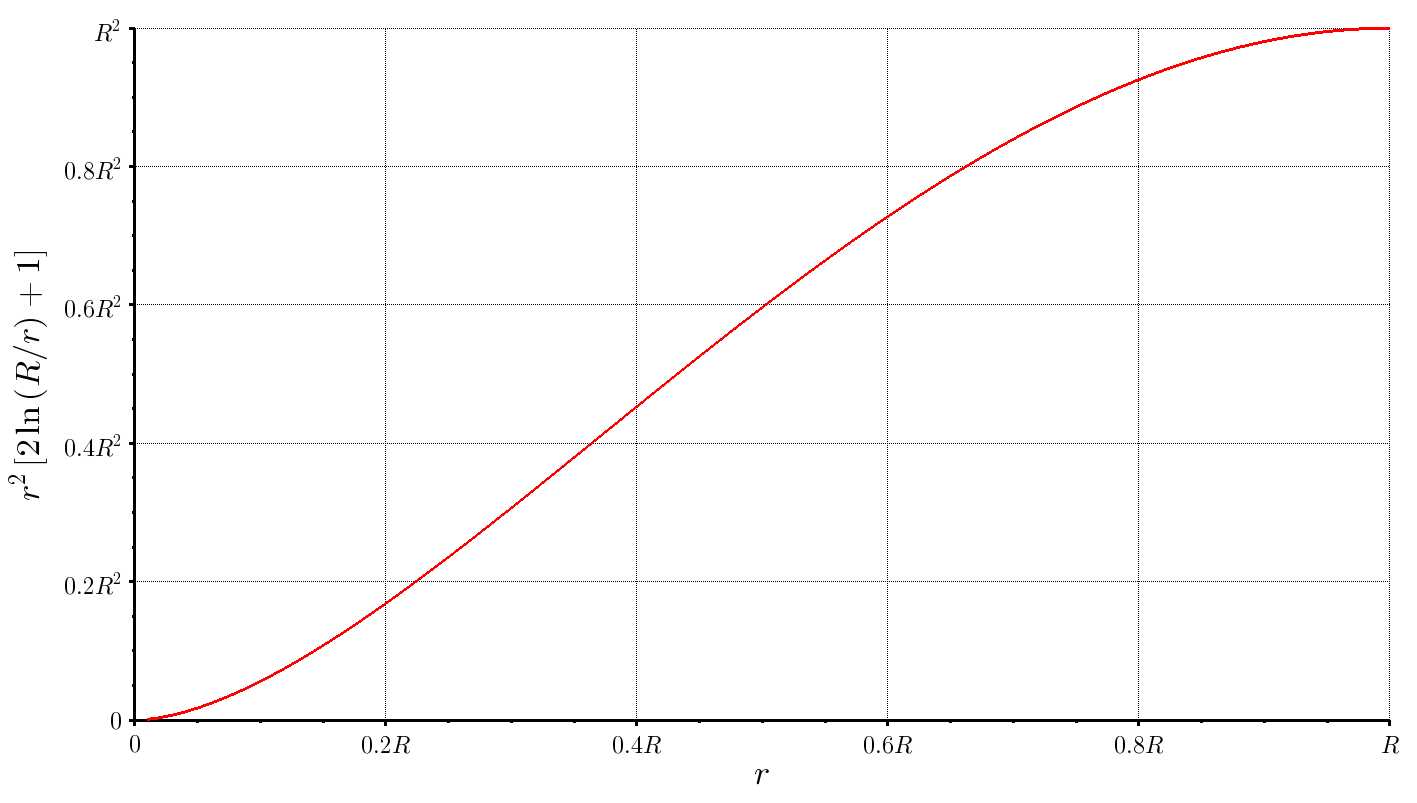}}
\caption{This plot reveals that the amplitude of the displacement current given by the last factor of Eq.~\ref{Id1} is bounded by $b^2$.}
\label{figcota}
\end{figure}

Although the amplitude of the displacement current is not constant we can find the maximum value in the region of interest. As shown in Fig.~\ref{figcota}, the last factor of Eq.~\ref{Id1}  is bounded by $b^{2}$, then
\begin{equation}
\label{Id2}
I_{D} \leq \frac{1}{4} \varepsilon_{0} \mu_{0} \omega^{2} I_{0} \cos (\omega t) b^{2}
\end{equation}
in the region $r<b$.

If the amplitude of the displacement current is much smaller than the amplitude of the conduction current $I_{D} \ll$ $I$, the relation given by Eq.~\ref{Id2}  can be expressed as
\begin{equation}
\label{Id3}
\frac{\varepsilon_{0} \mu_{0} \omega^{2} b^{2}}{4} \ll 1
\end{equation}

Given the speed of light is $c=1 / \sqrt{\varepsilon_{0} \mu_{0}}$, it can be rewritten as
\begin{equation}
\label{Id4}
\begin{split}
\frac{\omega^{2} b^{2}}{4 c^{2}} \ll 1 \\
\frac{b}{2 c} \ll \frac{1}{\omega} = \frac{T}{2 \pi}
\end{split}
\end{equation}
where $T$ is the period of the time-varying current  $I$ and 
consequently of the fields $\mathbf{E}$ and $\mathbf{B}$, while the quotient $b / c$ is the time $t$ that the fields needed to travel the distance $b$, that is,
\begin{equation}
\label{Id5}
t \ll \frac{T}{\pi}.
\end{equation}

We can derive an equivalent condition in terms of distances rather than time by revisiting Eq.~\ref{Id4} and expressing $\omega$ in term of the frequency $f$, resulting
\begin{equation}
\label{Id6}
\frac{2 \pi f b}{2 c} \ll 1,
\end{equation}
and
\begin{equation}
\label{Id7}
b \ll \lambda
\end{equation}
where $\lambda$ is the wavelength.

The inequality \ref{Id5} indicates that  $I_{D}$ is negligible when the time required for the fields to propagate from their source to the evaluation point is much shorter than the time scale over which the sources change state. This corresponds to a system in which the currents vary slowly. Similarly, the inequality \ref{Id7} shows that this condition holds when the system's dimensions are much smaller than the wavelength. Therefore, in systems where the currents vary slowly—or equivalently, where the system's dimensions are much smaller than the wavelength—Ampère’s law provides a good approximation for determining the magnetic field.

\section{\label{Comentarios}Implications for teaching}

In this article, we explored the role of displacement current in coaxial cables and resistors, highlighting  that it extends beyond charging capacitors to emerge as a fundamental aspect of time-varying electric fields. This broader perspective deepens students’ understanding of displacement current and its significance within electromagnetic theory.

The examples presented provide a valuable opportunity to explore the limits of validity of Ampère’s and Biot–Savart’s laws, fostering critical thinking and helping students develop physical intuition about the conditions under which certain effects—such as conduction currents—dominate over others, like displacement currents. Gaining insight into these distinctions is essential for the accurate application of electromagnetic principles to real-world systems.

From a teaching standpoint, the examples discussed in this paper provide physics instructors with a robust conceptual foundation for designing classroom sense-making activities aimed at fostering deeper understanding of displacement current and Ampère–Maxwell’s law. Sense-making activities \cite{hieggelke2015tipers, Maloney} engage students actively in evaluating and interpreting physical situations, challenging their conceptual understanding and encouraging critical thinking. Two particularly suitable examples are “What, if anything, is wrong?” tasks and "Ranking tasks". 
In the first case, instructors might use the resistor example to engage students in critical evaluation of the validity of common approximations, such as the Biot–Savart law. Alternatively, the coaxial cable scenario can be used to prompt students to reflect on the assumptions underlying the neglect of the displacement current term in the Ampère–Maxwell law—conditions under which the original form of Ampère’s law may serve as a valid approximation. "Ranking tasks" can also be constructed based on both examples, prompting students to compare and rank various scenarios according to the relative significance of displacement and conduction currents. These activities encourage students to critically analyze assumptions and approximations, thereby deepening their conceptual grasp of electromagnetic phenomena.

Both cases analyzed also serve as solid foundations for creating guided tutorials \cite{heron2004research} that systematically help students recognize and explore the specific physical and mathematical conditions determining the necessity or neglect of displacement current contributions. These instructional approaches encourage students to appreciate the coherent and coupled nature of Maxwell’s equations and support instructors in creating classroom environments that promote critical evaluation of common approximations used in electromagnetism.

Finally, by exploring the interplay between temporal and spatial scales, this work paves the way for more advanced topics. We believe that introducing these concepts— even qualitatively—at an introductory level fosters a deeper understanding of the transition from static to dynamic electromagnetic fields, preparing students for more advanced studies in electromagnetism.



\providecommand{\noopsort}[1]{}\providecommand{\singleletter}[1]{#1}%
\end{document}